\documentstyle[preprint,aps]{revtex}
\newcommand{\be}{\begin{equation}}
\newcommand{\ee}{\end{equation}}
\newcommand{\ba}{\begin{eqnarray}}
\newcommand{\ea}{\end{eqnarray}}
\begin{document}
\draft
\title{DYNAMICAL SUPERSYMMETRY OF A VORTEX SYSTEM WITH $1/r^{2}$ POTENTIAL}
\author{Younghoon Park and Yongsung Yoon}
\address{Department of Physics, Hanyang University, Seoul, 133-791 Korea}
\maketitle
\begin{abstract}
The Pauli Hamiltonian for a spin $\frac{1}{2}$ charged particle interacting
with a point magnetic vortex and  $1/r^{2}$ potential exhibits a dynamical
supersymmetry $Osp(1,1)$ on the plane except at the origin.
Using this symmetry, the spectrum and the wave functions have been obtained.
And, the dynamical supersymmetry could be imported to the case when an
external harmonic potential is added.
\end{abstract}


\narrowtext
\section{INTRODUCTION}

Several quantum mechanical systems can be completely and explicitly
solved with the aid of their dynamical symmetries.
The quantum mechanical system of a charged particle interacting with
a Dirac magnetic monopole possesses
a dynamical $O(2,1)$ symmetry \cite{Jackiwa}.
The Pauli Hamiltonian for a spin $\frac{1}{2}$ charged
particle in the presence of a magnetic monopole exhibits dynamical
supersymmetry $Osp(1,1)$ \cite{Hokera,Hokerb}.
Adding $1/r^{2}$ potential to the Pauli Hamiltonian,
the dynamical symmetry is enlarged to $Osp(2,1)$ \cite{Hokerb}.
The spectrum and wave functions of such systems can be constructed
by means of the group theoretical methods.

Recently, It has been shown that the quantum mechanical system of a charged
particle interacting
with a point magnetic vortex can be solved
using the dynamical symmetry $O(2,1)$ \cite{Jackiwb,Park,Demers}.
Interactions of charged particles with point magnetic vortices
in 2+1 dimensions frequently appear in physically interesting systems.
A typical example is a system of a charged particle interacting with
a long and thin solenoid.
We can describe such a system with the Hamiltonian for a charged
particle interacting with a point magnetic vortex in 2+1 dimensions.
The Pauli Hamiltonian of a spin $\frac{1}{2}$ charged particle
in the presence of a magnetic vortex
has dynamical supersymmetry $spl^{*}(2,1) $\cite{Park}.
Then, we can ask what the dynamical supersymmetry of a spin
$\frac{1}{2}$ charged particle interacting with a point magnetic vortex and
$1/r^{2}$ potential is.
We are going to realize the dynamical supersymmetry explicitly and to find
the spectrum and wave functions of this system
using the group theoretical methods.

\section{DYNAMICAL SUPERSYMMETRY OF THE SYSTEM}

In this section, we show that the Pauli Hamiltonian of a charged spin
$\frac{1}{2}$ particle interacting with a point magnetic vortex and
$1/r^{2}$ potential has dynamical supersymmetry $Osp(1,1)$.
The Pauli Hamiltonian in 2+1 dimensions which includes $1/r^{2}$
potential is given by
\be
H=\frac{1}{2m} (p_i - e A_i)^2 - \frac{e}{m} B
\sigma_{3} +\frac{\lambda^2}{2m r^2} + \frac{\lambda}{2m r^3}
\sigma_i r_i~,~~~~~ i=\{1,2\},
\ee
where $m$, $e$ and $\lambda$ are the particle's mass, charge and
a real parameter respectively. The vector potential  $A_i$ and
the corresponding magnetic field $B$ concentrated at the origin
are  externally given by
\ba
&&A_{i}({\bf r})=\frac{\Phi}{2 \pi} \epsilon^{ij} \frac{\hat{r}^j}{r}~,\\
&&B=\bigtriangledown \times {\bf A}=\Phi \delta^{(2)}({\bf r})~,
\ea
where $\Phi$ is the total magnetic flux.
The Hamiltonian $H$ can be derived from the Lagrangian
\be
L=\frac{1}{2} m {\dot r}_i^2 + e A_i {\dot r}_i -
\frac{\lambda^2}{2m r^2} + \frac{i}{2} \psi_\mu {\dot \psi}_\mu
+\frac{e}{m} B{\cal S} + \frac{i \lambda }{m r^3} \psi_i  \psi_{0} r_i~,~~~~~
\mu =\{0,1,2\}~,
\label{lag}
\ee
where ${\cal S}$ is the spin, ${\cal S}=-\frac{i}{2} \epsilon_{ij}
\psi_i \psi_j$.
$\{\psi_\mu\}$ are real generators of the Grassmann algebra,
which describe the spin degrees of freedom of a classical spin
$\frac{1}{2}$ particle \cite{Berezin}.
The rotationally invariant Lagrangian (\ref{lag}) gives
the following total angular moment $J$:
\be
J={\bf r} \times m {\bf v}  +  S  +  \nu~,
\ee
where $\nu = \frac{e\Phi}{2\pi}$.

It is well known that if a Lagrangian is changed by an exact differential
under an infinitesimal transformation,
then there exists a corresponding Noether charge.
Consider an infinitesimal transformation of the dynamical variables
$r_i$ and $\psi_\mu$
\be
r_i \rightarrow r_i + \delta r_i~,~~~~~
\psi_\mu \rightarrow \psi_\mu + \delta \psi_\mu~.
\label{transf}
\ee
If a Lagrangian is changed by a total time-derivative
\be
\delta L\;=\;\frac{d}{dt} \Lambda
\ee
under the transformation (\ref{transf}), the conserved Noether charge
$\hat{O}$ is
\be
\hat{O}=\delta r_i \frac{\partial L}{\partial {\dot r}_i}
+\delta \psi_\mu \frac{\partial L}{\partial {\dot \psi}_\mu}-\Lambda~.
\ee

The change of Lagrangian (\ref{lag}) under the transformation
(\ref{transf}) is
\ba
\delta L&=&\delta L_{1}+\delta L_{2}+\delta L_{3}~, \\
\delta L_{1}&=& \frac{1}{2} m {\dot r}_i \delta \dot r_i
+ \frac{i}{2}\delta \psi_\mu {\dot \psi}_{\mu}
- \frac{i}{2}\delta {\dot \psi}_\mu \psi_\mu~, \\
\delta L_{2}&=& e A_i \delta r_i
+ e \partial_i A_j {\dot r}_j \delta r_i
+ \frac{e}{m} \partial_i B \delta r_i {\cal S}
- i \frac{e}{m} B \delta \psi_i \epsilon_{ij} \psi_j~, \\
\delta L_{3}&=& \frac{\lambda^2 r_i}{m r^4} \delta r_i
+ \frac{i \lambda }{m} \partial_i (\frac {r_j}{r^3})
\delta r_i \psi_j \psi_{0}
+ \frac{i \lambda r_i }{m r^3} \delta \psi_i \psi_{0}
- \frac{i \lambda r_i }{m r^3} \delta \psi_{0} \psi_i~.
\ea
Therefore, under the super-transformations
\ba
\delta r_i&=&i \alpha f(t) \psi_i~,~~~
\delta \psi_i=\alpha m(-f(t) {\dot r}_i + \dot f(t) r_i )~,~~~
\delta \psi_{0}= \alpha f(t) \frac{\lambda}{r}~,
\ea
where $f(t)=\{1,t\}$ and $\alpha$ is a Grassmann super-transformation
parameter,
the Lagrangian (\ref{lag}) is changed by a total time-derivative
\ba
{\delta L}&=&\frac{d}{dt}(i \alpha f eA_i \psi_i
+i \frac{\alpha m f \psi_i {\dot r}_i}{2}
+i \frac{\alpha m \dot f \psi_i r_i}{2}
+i \frac{\alpha f \lambda \psi_{0}}{2 r})~.
\ea
And, the super-charges $Q$ and $S$ are given by
\ba
Q&=&\frac{1}{\sqrt m}(m \psi_i {\dot r}_i-\psi_{0}
\frac{\lambda}{r})~, \\
S&=&-tQ+\sqrt m \psi_i r_i~.
\ea

There are also dynamical bosonic symmetries in the Lagrangian (\ref{lag}):
\ba
\begin{array}{lll}
\delta_H r_i={\dot r}_i~,&\delta_H \psi_\mu= {\dot \psi}_\mu~,
&{\rm time~translation},  \\
\delta_D r_i=t {\dot r}_i-\frac{1}{2} r_i~,
&\delta_D \psi_\mu = t {\dot \psi}_\mu~,&{\rm dilation},\\
\delta_K r_i=t^2 {\dot r}_i-t r_i~,
&\delta_K \psi_\mu =t^2 {\dot \psi}_\mu~,
&{\rm conformal~transformation}.
\end{array}
\ea
The conserved bosonic charges corresponding to the bosonic symmetries are
\ba
H&=&\frac{m}{2} {\dot r}_i^2 - \frac{e}{m} BS
+\frac{\lambda^2}{2m r^2} - \frac{i \lambda}{m r^3} \psi_i \psi_{0} r_i~,
\nonumber  \\
D&=&tH - \frac{1}{4}m (r_i {\dot r}_i+{\dot r}_i r_i)~, \nonumber  \\
K&=&-t^2H+2tD+\frac{1}{2}m r^2~.
\ea

The canonical momentum  $p_i$ of the Lagrangian (\ref{lag}) is given by
\be
p_i=\frac{\partial L}{\partial {\dot r}_i} =m {\dot r}_i +eA_i~,
\ee
and the dynamical quantum algebra is
\ba
\begin{array}{lll}
&[r_i,r_j]=0~, &[r_i,{\dot r}_i]= i\delta_{ij}/m~, \\
&[{\dot r}_i,{\dot r}_j]=i \epsilon_{ij} eB/m^{2}~, &
\{ \psi_\mu,\psi_\nu \}=\delta_{\mu \nu}~.
\end{array}
\label{qu_al}
\ea
Because the canonical anti-commutation relations for $\psi_\mu$ define a real
Clifford algebra. $\psi_{\mu}$ can be given in terms of the Pauli matrices:
\be
\psi_0 = \frac{\sigma_{3}}{\sqrt 2}~,~~~~~
\psi_i = \epsilon _{ij} \frac{\sigma_j}{\sqrt 2}~,
\label{sig}
\ee
so that $S=\frac{1}{2} \sigma_{3}$ as expected.

The conserved charges $H$, $D, $K, $Q$, and $S$ satisfy the $Osp(1,1)$
super-algebra on the plane except at the origin \cite{Rittenberg,Nahm}:
\ba
\begin{array}{llll}
&[H,D]=iH~,    &    [H,K]=2iD~,    &       [D,K]=iK~,          \\
&\{Q,Q\}=2H~,  &    \{Q,S\}=-2D~,  &       \{S,S\}=2K~,        \\
&[H,Q]=0~,     &    [K,S]=0~,      &       [H,S]=-iQ~,         \\
&[K,Q]=iS~,    &    [D,Q]=-\frac{i}{2}Q,& [D,S]=\frac{i}{2}S~.
\end{array}
\ea
Because the all charges $H$, $D$, $K$, $Q$, and $S$ commute with $J$,
full symmetry of the system is
\be
G=SO(2)_{rotation} \; \otimes \; Osp(1,1)~.
\ee
The five charges are time-independent on the plane,
because their total time derivatives vanish.
For example, the total time derivative of the charge $K$ is
\be
\dot K=i[H,K]+\frac{\partial K}{\partial t}=0~.
\ee

To construct all the states, we introduce the Cartan basis for the
$Osp(1,1)$ algebra
\ba
R&=&\frac{1}{2 a^2} K+\frac{a^2}{2} H~, \\
B_\pm&=&\frac{1}{2 a^2} K-\frac{1}{2} a^2 H\pm iD~,  \\
F_\pm&=&\frac{1}{2a}S\mp i\frac{a}{2} Q~,
\ea
where $a$ is a real positive parameter and has dimension
$[t^\frac{1}{2}]$. $R$ is a compact generator, and
$B_\pm$ and $F_\pm$ are bosonic and fermionic raising/lowering operators
respectively. In this Cartan basis, $Osp(1,1)$ algebra becomes
\ba
\begin{array}{lll}
&[R,B_\pm]=\pm B_\pm~,  &[R,F_\pm]=\pm\frac{1}{2} F\pm~, \\
&[B_+ ,B_-]=-2R~,       &[B_\pm,F_\mp]=\mp F_\pm~,   \\
&\{F_+,F_-\}=R~,        &\{F_\pm,F_\pm\}=B_\pm~.
\label{comm}
\end{array}
\ea

The canonical chain for the group $Osp(1,1)$ is
\be
Osp(1,1)\;\supset\;O(2,1)\;\supset\;O(2)~,
\ee
and the corresponding Casimirs of each groups are $C$, $C_{0}$, and $R$.
The Casimir $C_{0}$ of the subgroup $O(2,1)$ is given by \cite{Book}
\be
C_{0}=\frac{1}{2} (HK + KH)-D^2~.
\ee
If we introduce the quantity  \(A=i[Q,S]-\frac{1}{2}\),
then $C_{0}$ is
\be
C_{0}=\frac{1}{4} A^2-\frac{A}{4}-\frac{3}{16}~.
\label{casa}
\ee
The Casimir of the super-group $Osp(1,1)$ is
\ba
C&=&\frac{1}{2} \{ H,K \}-D^2+\frac{i}{4}[Q,S]+\frac{1}{16}=\frac{1}{4} A^2~,
\label{casb}
\ea
which commutes with all the charges $H$, $D$, $K$, $Q$, and $S$ \cite{Hokerb}.
In the coordinate representation, the operator $A$ and $C$ are given by
\ba
A&=&J\sigma_3-\nu\sigma_3-\frac{\lambda}{r} r_i \sigma_i~,  \\
C&=&\frac{1}{4}(J^2+\nu^2-2J\nu+\lambda^2)~.
\ea

We can choose a mutually commuting set of operators $\{C,C_{0},R,J\}$,
which can label the quantum states completely.
However, $C$ and $C_{0}$ are not independent,
but can be expressed in terms of $A$ from the Eqs. (\ref{casa},\ref{casb}).
Moreover, the magnitude of the eigenvalue of the operator $A$ is not
independent of the angular momentum.
Therefore the eigenvalues of the following
three operators can specify the quantum states completely:
\be
\{J,sign(A),R\}~.
\ee

\section{SPECTRUM AND WAVE FUNCTIONS}

We are ready to determine the spectrum and the wave functions using algebraic
methods. To do so, we must know the eigenvalues of the three operators
$J$, $sign(A)$, and $R$, which can label the quantum states.
The eigenvalue of $J$ is denoted by $j$.
The eigenvalue of $R$ is $(\Delta_{j,\alpha} +n)$, where $\Delta_{j,\alpha}$
is related to the eigenvalue of $C_{0}$:
\be
C_{0}\;=\;\Delta_{j,\alpha} ( \Delta_{j,\alpha} - 1\;)~,
\ee
and $\alpha=\pm 1$ denotes the eigenvalue of $sign(A)$.
Therefore, the expressions for the quantum numbers are
\ba
J \mid j,\alpha,n > &=& j \mid j,\alpha,n >~,   \\
A \mid j,\alpha,n > &=& \alpha D_j \mid j,\alpha,n >~, \label{A_eq} \\
R \mid j,\alpha,n > &=& (\Delta_{j,\alpha}+\;n) \mid j,\alpha,n >~,
\ea
where
\ba
D_j &=& \sqrt{j^2 + \nu^2 -2j \nu + \lambda^2}~, \\
\Delta_{j,\alpha} &=& \frac{1}{2} \mid D_j -\frac{\alpha}{2}\mid+\frac{1}{2}~.
\ea
$\Delta_{j,\alpha}$ can be expressed in terms of $D_{j}$ using the Eq.
(\ref{casa}):
\be
\Delta_{j,\alpha} ( \Delta_{j,\alpha} - 1)=
\frac{1}{4} D_j^2-\frac{1}{4}D_j-\frac{3}{16}~.
\ee
{}From the commutation relations (\ref{comm}), we obtain
\ba
B_\pm \mid j,\alpha,n > &=& \sqrt{(\Delta_{j,\alpha}+n)(\Delta_{j,\alpha}
+n\pm 1)-\Delta_{j,\alpha}(\Delta_{j,\alpha}-1) }\;\mid j,\alpha,n\pm 1>~,\\
F_\pm \mid j,\alpha,n > &=& \sqrt{\frac{1}{2} (\Delta_{j,\alpha}+n) \pm
\frac{1}{8} \pm \frac{1}{4} \alpha D_j } \;\mid j,-\alpha,n-\frac{1}{2}
\alpha \pm \frac{1}{2}>~.
\ea

The lowest states can be obtained by the first order differential
equation
\be
B_- \mid j,\alpha,0>\;=\;0~.
\ee
{}From this equation, we have two lowest states
$\mid j,1,0>$ and $\mid j,-1,0>$.
The higher states can be constructed by applying $F_+$
and $B_+$ on the two lowest states,
\ba
\mid j,\alpha,n>&=&
\sqrt{\frac{\Gamma(2 \Delta_{j,\alpha})}
{n!\;\Gamma(2 \Delta_{j,\alpha} + n) }} \; B_+^n \mid j,\alpha,0>~,  \\
\mid j,-1,n>&=&
\frac{1}{\sqrt{\frac{1}{2}D_j+\frac{1}{4}+\frac{n}{2}}} \; F_+ \mid j,1,n>.
\ea
Because all the quantum states have been constructed,
the corresponding wave functions can be easily obtained.
The angular part of the wave function is found by obtaining
the eigenvectors of the operator $A$.
Assuming the normalized eigenvectors of $A$ as
\be
\mid j,\alpha>\;=\;C^{(\alpha)}_+ \mid j-\frac{1}{2}> \otimes \mid
\frac{1}{2}>\;+\;C^{(\alpha)}_- \mid j+\frac{1}{2}> \otimes \mid
-\frac{1}{2}>~,
\ee
the angular eigenfunction can be given in the form
\ba
\eta_{j,\alpha}(\theta) &=& <\theta \mid j,\alpha>   \nonumber \\
&=& \left( \begin{array}{c}
           C^{(\alpha)}_+e^{i(j-\frac{1}{2}) \theta} \\
           C^{(\alpha)}_-e^{i(j+\frac{1}{2}) \theta}
           \end{array}  \right)~.
\ea
To determine $C^{(\alpha)}_\pm$, we use the Eq. (\ref{A_eq}) on the states
\ba
A \left( \begin{array}{c} C^{(\alpha)}_+e^{i(j-\frac{1}{2}) \theta} \\
          C^{(\alpha)}_-e^{i(j+\frac{1}{2}) \theta} \end{array}  \right)
&=& \left( \begin{array}{ll} j-\nu \;\;\;\;\;&-\lambda e^{-i\theta} \\
           -\lambda e^{i \theta} \;\;\;\;\;&-j+\nu \end{array} \right)
    \left( \begin{array}{c} C^{(\alpha)}_+e^{i(j-\frac{1}{2}) \theta} \\
    C^{(\alpha)}_-e^{i(j+\frac{1}{2}) \theta} \end{array}  \right)
    \nonumber  \\
&=& \alpha D_j \left( \begin{array}{c}
          C^{(\alpha)}_+e^{i(j-\frac{1}{2}) \theta} \\
          C^{(\alpha)}_-e^{i(j+\frac{1}{2}) \theta} \end{array} \right)~.
\ea
The ratio of $C^{(\alpha)}_+$ to $C^{(\alpha)}_-$ is
\be
\frac{C^{(\alpha)}_+}{C^{(\alpha)}_-}=
\frac{j-\nu+\alpha D_j}{-\lambda}=\frac{\lambda}{j-\nu-\alpha D_j}~.
\ee
Therefore, the angular parts of the wave functions are
\be
\eta_{j,\alpha}(\theta)=N_{j,\alpha}
     \left( \begin{array}{c}(j-\nu+\alpha D_j) e^{i(j-\frac{1}{2}) \theta} \\
     -\lambda e^{i(j+\frac{1}{2}) \theta} \end{array}  \right)~,
\label{eta}
\ee
where $N_{j,\alpha}$ is the normalization factor.

The radial wave function $\Psi_{j,\alpha,n}(r)$ is constructed with
the help of the identity;
\be
R\;=\;\frac{1}{2 a^2} K + \frac{a^2}{2K} (D^2-iD+\frac{1}{4}A^2-\frac{1}{4}A
-\frac{3}{16})~.
\ee
The eigenvalue equation for $R$ in the coordinate representation is
\be
<r,\theta \mid R \mid j,\alpha,n>\;=\;(\Delta_{j,\alpha} + n)
<r,\theta \mid j,\alpha,n>~.
\ee
The full wave function can be written by
\be
<r,\theta \mid j,\alpha,n>\;=\;\Psi_{j,\alpha,n}(r)\;\eta_{j,\alpha}(\theta)~.
\ee
Because the charges are time-independent, we can use the coordinate
representations of $K$ and $D$ at $t=0$:
\be
K=\frac{1}{2}m r^2~,~~~~~
D=\frac{i}{2}(r \frac{\partial}{\partial r} + 1)~.
\ee
Therefore, the differential equation for $\Psi_{j,\alpha,n}(r)$ is
\be
\frac{1}{2m}(-\frac{1}{r} \frac{d}{dr}-\frac{d^2}{dr^2}
+\frac{{(2\Delta_{j,\alpha}-1)}^2}
{r^2}+\frac{m^2}{a^4}r^2)\Psi_{j,\alpha,n}(r)=\frac{2}{a^2}
(\Delta_{j,\alpha} + n) \Psi_{j,\alpha,n}(r)~.
\ee
We find the solutions of the above equation in terms of the generalized
Laguerre polynomials,
\be
\Psi_{j,\alpha,n}(r)={(-1)}^n {(\frac{m}{a^2})}^{\Delta_{j,\alpha}}
\sqrt{(\frac{n!}{\pi\Gamma(2 \Delta_{j,\alpha} + n) })}
\; r^{2 \Delta_{j,\alpha}-1} {\rm exp}(-\frac{mr^2}{2a^2})
L^{2 \Delta_{j,\alpha}-1}_n (\frac{mr^2}{a^2})~.
\ee
Of course, it is possible to obtain the above wave functions successively
applying the coordinate representations of $B_{+}$ to the ground state
wave function $<r,\theta \mid j,\alpha,0>$.

Also, the energy eigenfunctions can be obtained using the following identity;
\be
H\;=\;\frac{1}{K} ( D^2 - iD + \frac{A^2}{4}-\frac{A}{4}-\frac{3}{16})~.
\ee
The radial Schr\"{o}dinger equation is
\be
\frac{1}{2m} ( -\frac{1}{r} \frac{d}{dr}\;-\; \frac{d^2}{dr^2}\;+\;
\frac{{(2\Delta_{j,\alpha}-1)}^2}{r^2} ) \Psi_{j,\alpha}^E (r)\;=\;
E \Psi_{j,\alpha}^E (r)~.
\ee
The regular solutions to the above equation are
\be
\Psi_{j,\alpha}^E (r)\;=\;\sqrt m J_{2\Delta_{j,\alpha}-1}(\sqrt {2mE}\;r)~.
\ee

All the regular scattering wave functions vanish at the origin and
respect the dynamical $Osp(1,1)$ symmetry.
However, if $(2\Delta_{j,\alpha} -1) < 1$, i.e.,
\be
\nu - \sqrt{\frac{5}{4}+\alpha-\lambda^{2}} < j <
\nu + \sqrt{\frac{5}{4}+\alpha-\lambda^{2}}~,
\label{limit}
\ee
then there exists one parameter family of square integrable self-adjoint
singular wave functions which do not respect the dynamical symmetry at all:
\be
\Psi_{j,\alpha}^E (r)\;=\;N (
J_{2\Delta_{j,\alpha}-1}(\sqrt {2mE}\;r)
+ \xi J_{-2\Delta_{j,\alpha}+1}(\sqrt {2mE}\;r))~,
\ee
where $N$ and $\xi$ are the normalization and the real self-adjoint extension
parameter respectively \cite{Jackiwc}.
It is found that the number of these singular solutions is zero or two
for an integer $\nu$, one or three for a half integer $\nu$,
and at most three for a general $\nu$ from the Eq. (\ref{limit}).

\section{The case when an external harmonic potential is added}
In this section, we show that the dynamical supersymmetry $Osp(1,1)$ of
a spin $\frac{1}{2}$ charged particle interacting with a point
vortex and $1/r^2$ potential can be imported to
the case when an external harmonic oscillator potential is added.

We transform coordinates by
\ba
\begin{array}{c}
t=\omega^{-1} tan (\omega t^\prime)~,  \\
r_i=r^\prime_i sec(\omega t^\prime)~, \\
\psi_\mu=\psi^\prime_\mu~.
\end{array}
\ea
Then, the action for Lagrangian,
\be
S=\int dt ( \frac{1}{2} m {\dot r}_i^2 +e A_i {\dot r}_i
-\frac{\lambda^2}{2m r^2} + \frac{i}{2} \psi_\mu {\dot \psi}_\mu
+\frac{e}{m} B{\cal S} + \frac{i \lambda }{m r^3} \psi_i \psi_{0} r_i)~,
\ee
is transformed into $S^h=\int dt^\prime L^h$.
The transformed ${\dot r}_i$, $A_i$, and $B$ are
\ba
\begin{array}{ll}
{\dot r}_i= {\dot r}^\prime_i cos(\omega t^\prime)
+ \omega r^\prime_i sin(\omega t^\prime)~,
&A_i=A^\prime_i cos (\omega t^\prime)~,  \\
B=B^\prime cos^2(\omega t^\prime)~,
&{\dot \psi}_\mu={\dot \psi}^\prime_\mu cos^2(\omega t^\prime)~.
\end{array}
\ea
The new Lagrangian is
\be
L^h=\frac{1}{2} m {\dot r}^{\prime 2}_i-\frac{1}{2} m \omega^2 r^{\prime 2}
+e A^\prime_i {\dot r}^\prime_i -\frac{\lambda^2}{2m r^{\prime 2}}
+ \frac{i}{2} \psi^\prime_\mu {\dot \psi}^\prime_\mu
+\frac{e}{m} B^\prime{\cal S}^\prime
+ \frac{i \lambda }{m r^{\prime 3}} \psi^\prime_i \psi^\prime_{0} r^\prime_i ~,
\ee
where ${\dot \psi}_i= \frac{d}{dt^\prime} \psi, ~{\dot r}^\prime_i=
\frac{d}{dt^\prime}r^\prime_i$.
And $A^\prime_i$ and $B^\prime$ are  as the unchanged,
but in terms of $r^\prime_i$.

The corresponding new Hamiltonian is
\be
{\sf H}^h=\frac{1}{2} m {\dot r}_i + \frac{1}{2} m \omega^2 r^2
+\frac{\lambda^2}{2m r^2} -\frac{e}{m} B {\cal S}
-\frac{i \lambda }{m r^3} \psi_i \psi_{0} r_i~,
\ee
where the primes are suppressed, and $m {\dot r}_i$ is $p_i-eA_i$.
This Hamiltonian describes a spin $\frac{1}{2}$ particle interacting
with a point vortex and a $1/r^2$ potential, but with the external
harmonic oscillator potential added.
Using the relation of the change of coordinate, we can obtain the
imported charges
\ba
H^h&=&\frac{1}{2}m{\dot r}_i^2 cos^2(\omega t)+\frac{1}{4} m\omega
(r_i {\dot r}_i+{\dot r}_i r_i ) sin(2\omega t)
+\frac{1}{2}m\omega^2 r^2 sin^2(\omega t)   \nonumber \\
&&-\frac{e}{m} B{\cal S} cos^2(\omega t)
+cos^2(\omega t) \frac{\lambda^2}{2mr^2}
-cos^2(\omega t) \frac{i \lambda }{m r^3} \psi_i \psi_{0} r_i~,  \nonumber \\
D^h&=&\frac{1}{\omega}tan(\omega t)H
-\frac{1}{4}m (r_i {\dot r}_i+{\dot r}_i r_i)
-\frac{1}{2}m \omega r^2 tan(\omega t)~,   \nonumber \\
K^h&=&-\frac{1}{\omega^2}tan^2(\omega t)H^h+\frac{2}{\omega}tan(\omega t)D^h
+\frac{1}{2}m r^2 sec^2(\omega t)~,  \\
Q^h&=&\sqrt m cos(\omega t) \psi_i {\dot r}_i
+\sqrt m \omega sin(\omega t) \psi_i r_i
-cos(\omega t) \frac{\lambda}{\sqrt m r} \psi_0~, \nonumber \\
S^h&=&-\frac{1}{\omega}tan(\omega t)Q^h+\sqrt m sec(\omega t) \psi_i r_i~.
\nonumber
\ea
These charges are conserved with respect to the new Hamiltonian
${\sf H}^h$. From Eqs. (\ref{qu_al},\ref{sig}), we can check they satisfy
the supersymmetry $Osp(1,1)$.

Using the imported charges, we get at $t=0$:
\ba
&&{\sf H}^h=\omega^2 K^h+H^h=2\omega R^h \mid_{a^2=\omega^{-1}}~,  \\
&&R^h=\frac{1}{2a^2}K^H+\frac{a^2}{2}H^h~. \nonumber
\ea
The eigenvalue equation for $R^h$ is
\be
R^h \mid j,\alpha,n > = (\Delta_{j,\alpha}+\;n) \mid j,\alpha,n >~.
\ee
Thus, the energy spectrum of ${\sf H}^h$ is given by
\be
E^h=2(\Delta_{j,\alpha}+n)\omega ~.
\ee
The radial energy eigenfunctions are found using the identity for $R^h$;
\be
R^h\;=\;\frac{1}{2 a^2} K^h + \frac{a^2}{2K^h} ((D^h)^2-iD^h+C_0)~.
\ee
Because the Casimir of $C_0$ is $\Delta_{j,\alpha}(\Delta_{j,\alpha}-1)$,
and the coordinate representation of $D^h$ at $t=0$ is
$\frac{i}{2}(r \frac{\partial}{\partial r}+1)$, the eigenvalue equation
for $R^h$ in the coordinate representation becomes
\be
\frac{1}{2m}(-\frac{1}{r} \frac{d}{dr}-\frac{d^2}{dr^2}
+\frac{{(2\Delta_{j,\alpha}-1)}^2}{r^2}+m^2\omega^2 r^2)\Psi^h_{j,\alpha,n}(r)
=2\omega(\Delta_{j,\alpha} + n) \Psi^h_{j,\alpha,n}(r)~.
\ee
Therefore, the radial wavefunctions are
\be
\Psi^h_{j,\alpha,n}(r)={(-1)}^n {(m\omega)}^{\Delta_{j,\alpha}}
\sqrt{(\frac{n!}{\pi\Gamma(2 \Delta_{j,\alpha} + n) })}
\; r^{2 \Delta_{j,\alpha}-1} {\rm exp}(-\frac{1}{2}m \omega r^2)
L^{2 \Delta_{j,\alpha}-1}_n (m\omega r^2)~.
\ee
The angular part of the wavefunction is the same as Eq. (\ref{eta}).

\section{conclusion}
We have shown that the quantum mechanical system of a spin 1/2 charged particle
interacting with a point magnetic vortex and $1/r^{2}$ potential possesses the
dynamical symmetry $Osp(1,1)$. Using this dynamical symmetry, the spectrum
and wave functions of the system could be constructed. And, the dynamical
supersymmetry could be imported to the case when additional harmonic potential
is present.

 ~\\
\noindent
{\it Acknowledgments:}
This work was supported in part by the Ministry of Education through Grant No.
BSRI-2441 and the Korea Science and Engineering Foundation(through Grant No.
94-1400-04-01-3 and CTP at SNU).

\end{document}